\documentclass[12pt]{article}
\usepackage{epsfig}
\usepackage{epsfig}
\begin{document}
\def\tr{{\rm tr}\, }
\def\Tr{{\rm Tr}\, }
\def\hTr{\hat{\rm T}{\rm r}\, }
\def\be{\begin{eqnarray}}
\def\ee{\end{eqnarray}}
\def\ctt{\chi_{\tau\tau}}
\def\cta{\chi_{\tau a}}
\def\ctb{\chi_{\tau b}}
\def\cab{\chi_{ab}}
\def\cba{\chi_{ba}}
\def\ptt{\phi_{\tau\tau}}
\def\pta{\phi_{\tau a}}
\def\ptb{\phi_{\tau b}}
\def\>{\rangle}
\def\<{\langle}
\def\d{\hbox{d}}
\def\pab{\phi_{ab}}
\def\lb{\label}
\def\appendix{{\newpage\section*{Appendix}}\let\appendix\section%
        {\setcounter{section}{0}
        \gdef\thesection{\Alph{section}}}\section}
\renewcommand{\figurename}{Fig.}
\renewcommand\theequation{\thesection.\arabic{equation}}
\hfill{\tt IUB-TH-057}\\\mbox{}
\vskip0.3truecm
\begin{center}
\vskip 2truecm {\Large\bf Holography with Gravitational Chern-Simons Term}
\vskip 1.5truecm
{\large\bf
Sergey N.~Solodukhin\footnote{
{\tt s.solodukhin@iu-bremen.de}}
}\\
\vskip 0.6truecm
\it{School of Engineering and Science, \\
International University Bremen, \\
P.O. Box 750561,
Bremen 28759,
Germany}
\end{center}
\vskip 1cm
\begin{abstract}
\noindent
The holographic description in the presence of  gravitational Chern-Simons term
is studied. The modified gravitational equations are integrated by using the Fefferman-Graham expansion and the
holographic stress-energy tensor is identified. The stress-energy tensor has both conformal anomaly and gravitational
or, if re-formulated in terms of the zweibein,  Lorentz anomaly. We comment on the structure of
anomalies in two dimensions and show that the two-dimensional stress-energy tensor can be reproduced
 by integrating the
conformal and gravitational anomalies. We study the black hole entropy in theories with a gravitational Chern-Simons term
and find that the usual Bekenstein-Hawking entropy is modified. For the BTZ black hole the modification
is determined by  area of  the inner horizon. We show that the total entropy of the BTZ black hole is precisely reproduced
in a boundary CFT calculation using the Cardy formula.
\end{abstract}
\vskip 1cm
\newpage

\section{Introduction}
\setcounter{equation}0
It is amazing how much of physics  is encoded in  geometry of asymptotically
anti-de Sitter (AdS) space-time. This includes the information on the ultra-violet
divergences, quantum effective action and the conformal anomaly. The latter is important element in the
holographic description and is due to peculiar  nature of the asymptotic  diffeomorphisms that generate
conformal symmetry \cite{Brown:1986nw}, \cite{FG}. The gravitational Einstein-Hilbert action,
in which metric  is  fixed on the boundary,
breaks the asymptotic conformal symmetry and is thus the source of the anomalies.
This and other related issues were much studied
in recent years \cite{Henningson:1998gx}-\cite{BT}.

In recent interesting paper \cite{Kraus} Kraus and Larsen  have modified the gravitational action in three dimensions
by adding the
gravitational Chern-Simons term. The resultant theory is known as the topologically massive gravity
\cite{Deser1}, \cite{Deser2}. Appearance of the Chern-Simons terms is generically predicted in string theory.
The gravitational Chern-Simons term, explicitly depending on connection, is gauge invariant only
up to some boundary terms. So that its presence necessarily breaks the asymptotic
coordinate invariance. The appearance of the gravitational anomaly\footnote{For review on the gravitational anomalies see
the original works \cite{Alvarez-Gaume:1983ig}, \cite{Bardeen:1984pm}.}
in  the boundary  theory is thus should be expected
in addition to the already existing conformal anomaly.
Alternatively, if the Chern-Simons term is defined in terms of the Lorentz connection, the asymptotic local
Lorentz symmetry is broken and the Lorentz anomaly should appear.
All these expectations were explicitly confirmed in \cite{Kraus} by looking at  how the gravitational
action changes under the gauge transformations. On the boundary side the anomaly arises due to different
central charge in  holomorphic and anti-holomorphic sectors. Such theories were  studied some time ago, see in particular
 \cite{Leutwyler:1984nd},
\cite{Myers:1992ea}.

In the present note, inspired by \cite{Kraus}, we take a different route to anomalies
and show that they follow directly  from the bulk gravitational equations. The latter are integrated
by expanding the bulk metric in powers of distance from the boundary.
This is much in  the spirit of \cite{FG}, \cite{SS}, \cite{CMP}.
The integration procedure involves the fixing of the boundary data. The data are the boundary metric and the
holographic stress-energy tensor.
This helps to determine explicitly the structure of
stress-energy tensor in terms of the coefficients in the expansion and
completely fix the form of the anomalies. The gravitational anomaly we get agrees with the one obtained in \cite{Alvarez-Gaume:1983ig}
and \cite{Leutwyler:1984nd}.
In the dual picture the holographic stress-energy tensor  should be identified with the quasi-local stress-energy tensor
which determines the values of mass and angular  momentum.

The gravitational Chern-Simons term is eligible gravitational action
which produces covariant equations of motion that are, in particular, solved by the BTZ metric.
It is therefore interesting question whether the black hole entropy is
modified when the Chern-Simons  term is included. We study this question and obtain a
contribution to the entropy due to the gravitational Chern-Simons term. This contribution depends on the
value of the Lorentz connection at the horizon and is, nevertheless,  gauge invariant.
For the BTZ black hole the entropy due to Chern-Simons term is proportional to  the area of  inner horizon.
This is surprising, taking that in any  theory non-linear in Riemann curvature the entropy of BTZ black hole
 is always, as we argue in the paper, determined by  area of the outer horizon.
 In the theory at hand, the total entropy has dual meaning in terms of the
boundary CFT and is precisely reproduced by means of the Cardy formula, as we show.

\section{Fefferman-Graham expansion and gravitational \\
anomaly}
\setcounter{equation}0
The gravitational theory on three-dimensional space-time is given by the action
\be
I_{\tt gr}=I_{\tt EH}+I_{\tt CS}~~,
\lb{1}
\ee
which is sum of ordinary Einstein-Hilbert action (with cosmological constant)
\be
I_{\tt gr}=-{1\over 16 \pi G_N}[\int_M (R[G]+2/l^2)+\int_{\partial M}2K],
\lb{2}
\ee
where $K$ is trace of the second fundamental form of boundary $\partial M$,
and the gravitational Chern-Simons term
\be
I_{\tt CS}={\beta \over 64\pi G_N} \int_Mdx^3 \epsilon^{\mu\nu\alpha} [R_{ab\mu\nu}\omega^{ab}_{\ \ ,\alpha}+{2\over 3}
\omega^{a}_{\ b,\mu}\omega^b_{\ c,\nu}\omega^c_{\ a,\alpha}]~~.
\lb{3}
\ee
Parameter $l$ in (\ref{2}) sets the AdS scale. To simplify things, on intermediate stage of calculation, we take liberty
to  use units $l=1$ restoring $l$ explicitly in the final expressions.
Parameter $\beta$ has dimension of length.
We use the following definition for the curvature $R^a_{\ b\mu\nu}\equiv \partial_\mu\omega^a_{\ b,\nu}+
\omega^a_{\ c,\mu}\omega^c_{\ b,\nu}-(\mu \leftrightarrow \nu)$. The torsion-free Lorentz connection
$\omega^{a}_{\ b}=\omega^a_{\ b,\mu}dx^\mu$ is determined as usual by equation
\be
de^a+\omega^a_{\ b}\wedge e^b=0~~,
\lb{4}
\ee
where the orthonormal basis $e^a=h^a_\mu dx^\mu~,a=1,2,3$ is "square root" of the metric,
$G_{\mu\nu}=h^a_\mu h^b_\mu\eta_{ab}$. The equation (\ref{4}) can be used to express
components of the Lorentz connection in terms of $h^a_\mu$ and their derivatives\footnote{We use
the Latin letters $(a,b,c,d)$  for the inner Lorentz indices and Greek letters $(\mu,\nu,\alpha,\beta..)$
for the coordinate indices.}
\be
&&\omega_{ab,\mu}={1\over 2}(C_{a\nu\mu}h^\nu_b+C_{b\mu\nu}h^\nu_a-C_{d\alpha\beta}h^\alpha_ah^\beta_bh^d_\mu)~~, \nonumber \\
&&C^a_{\mu\nu}\equiv \partial_\mu h^a_\nu-\partial_\nu h^a_\mu~~.
\lb{L}
\ee
The Levi-Civita symbol is determined as $\epsilon^{\mu\nu\alpha}=h_a^\mu h_b^\nu h_c^\alpha
\epsilon^{abc}$. To complete our brief diving into theory of gravity in orthonormal basis we remind the reader
that $h^a_\mu$ is covariantly constant,
\be
\nabla_\mu h^a_\nu=\partial_\mu h^a_\nu -\Gamma^\lambda_{\mu\nu} h^a_\lambda+\omega^a_{\ b,\mu}h^b_\nu=0~~,
\lb{h}
\ee
that is of course equivalent to the equation ({\ref{4}). The latter property is useful in that we may freely manipulate
with $h^a_\mu$ by pulling it inside the covariant derivative or taking  it out. This property also means that the Levi-Civita
symbol  is covariantly constant, $\nabla_\sigma\epsilon^{\mu\nu\alpha}=0$.

The theory described by the action (\ref{1}) is quite well known and belongs to the class of theories with topological
mass \cite{Deser1}, \cite{Deser2}. A remarkable property of this theory is that it describes a propagating degree of freedom
although each term in (\ref{1}) taken separately is topological and thus does not contain local degrees of freedom.
Another interesting property of (\ref{1}) is that the gravitational Chern-Simons  tem explicitly breaks the
asymptotic coordinate invariance if expressed in terms of the metric connection $\Gamma^\alpha_{\ \beta \mu}$
or the asymptotic local Lorentz invariance if the term is written using the Lorentz connection as in (\ref{3}).
The violation of the gauge symmetry happens only asymptotically because the variation of the Chern-Simons term under
local gauge symmetry generates terms on the boundary of the space-time. This violation thus should be manifest
in the boundary theory. Indeed, in \cite{Kraus} this was related to the appearance of the gravitational or Lorentz
anomalies in the boundary theory. Such anomalies are natural when $c_L \neq c_R$ in the boundary conformal field theory.
Such theories were studied  some time ago, see for example \cite{Myers:1992ea}.   In the present context,
these anomalies are obtained holographically and are encoded in the dynamics of the gravitational field in the bulk.
In \cite{Kraus} the anomalies were derived by looking at how the gravitational action (\ref{1}) changes under
the gauge symmetries. Here we look at the problem at somewhat different angle. We demonstrate that  anomalies show up in
the process of the holographic reconstruction of the bulk metric from the boundary data. In the absence of the gravitational
Chern-Simons term the bulk metric is uniquely determined once the holographic boundary data, the boundary metric
representing  the conformal class and the  boundary stress tensor, are specified. Usually the boundary stress tensor
is not entirely arbitrary. It is covariantly conserved and its trace should reproduce the conformal anomaly.
The anomaly itself is completely specified by the boundary metric. The details of this analysis
can be found in \cite{Henningson:1998gx},
\cite{SS}, \cite{CMP}. The presence of the gravitational Chern-Simons term in the bulk action manifests
itself in an interesting way: the boundary stress tensor is no more covariantly conserved. This is how the gravitational anomaly
in the boundary theory shows up. In order to see this explicitly we solve the gravitational bulk equations
modified by the presence of the Chern-Simons term starting from the boundary and finding the bulk metric as an expansion,
well-known in the physics and mathematics literature as the Fefferman-Graham expansion \cite{FG}.

The gravitational bulk equations obtained by varying the action (\ref{1}) with respect to metric
takes the form
\be
R_{\mu\nu}-{1\over 2}G_{\mu\nu}R-G_{\mu\nu}+\beta C_{\mu\nu}=0~~,
\lb{*}
\ee
where all curvature tensors are determined with respect to the bulk metric $G_{\mu\nu}$. The tensor $C_{\mu\nu}$  is result
of the variation of the of the gravitational Chern-Simons term. It is known as the Cotton tensor and takes the form
\be
C_{\mu\nu}=\epsilon_\mu^{\ \alpha\beta}\nabla_\alpha(R_{\beta\nu}-{1\over 4} G_{\beta\nu}R)~~.
\lb{C}
\ee
Although the Chern-Simons (\ref{3}) is defined in terms of the Lorentz connection which is not gauge invariant object
its variation is presented in the covariant and gauge invariant form (\ref{C}). This is just a manifestation of the fact
that the "non-invariance" of the Chern-Simons term resides on the boundary and does not appear in the bulk field equations.
By virtue of the Bianchi identities this quantity is symmetric, manifestly traceless and identically
covariantly conserved,
$$
C_{\mu\nu}G^{\mu\nu}=0~,~~\nabla_\mu C^{\mu}_{\ \nu}=0~,~~\epsilon_{\alpha\mu\nu}C^{\mu\nu}=0~~.
$$
Due to  these properties we find that solution to the equation (\ref{*}) is space-time with
constant Ricci scalar $R=-6$. This is exactly what we had when the Chern-Simons term was not included in the action.
In that case we had moreover that $R_{\mu\nu}=-2G_{\mu\nu}$ and the solution was the constant curvature space.
It is no more the case in the presence of the Chern-Simons term and we have
\be
R_{\mu\nu}=-2G_{\mu\nu}-\beta C_{\mu\nu}~~.
\lb{5}
\ee
This is that equation which we are going to solve.
We start with choosing the bulk metric in the form
\be
ds^2=G_{\mu\nu}dX^\mu dX^\nu=dr^2+g_{ij}(r,x)dx^idx^j
\lb{6}
\ee
that always can be done by using appropriate normal coordinates.
The quantity $g_{ij}(r,x)$ is induced metric on the hypersurface of constant radial coordinate $r$.
The following expansion
\be
g(r,x)=e^{2r}[g_{(0)}+g_{(2)}e^{-2r}+g_{(4)}e^{-4r}+...]
\lb{7}
\ee
is  assumed
so that the metric (\ref{6}) describes asymptotically anti-de Sitter space-time with $g_{(0)}$ being the metric on its
two-dimensional boundary. In the case of pure GR, described by action (\ref{2}), the solution to the gravitational
equations contains
\cite{SS} only these three terms in the expansion (\ref{7}). This is no more true when the Chern-Simons term is turned on
and the whole infinite series should be expected in (\ref{7}). The presence of infinite number of terms in the expansion
(\ref{7}) generically seems to be related to the presence of local propagating degrees of freedom in the theory.

Now the routine is to insert the expansion (\ref{7}) into the gravitational equations (\ref{5})
and equate the coefficients appearing in front of the same power of $e^r$ on both sides of the equation.
This gives certain constraints on the coefficients $g_{(2n)}$ appearing in the expansion (\ref{7}) allowing express
$g_{(2n)}$ in terms of the coefficients $g_{(2k)}$ with $k<n$. Generically, in odd dimension $(d+1)$ there may appear also
"logarithmic" term $h_{(d)}r e^{-dr}$ in (\ref{7}). $h_{(d)}$ is traceless and covariantly conserved and is
local function of boundary metric $g_{(0)}$. In  $d=2$  no such local function of two-dimensional metric exists
so that $h_{(2)}$ identically vanishes (see \cite{SS}, \cite{CMP}). When gravitational Chern-Simons term
is present same arguments are valid so that no logarithmic term is likely to appear. In any event
it would not affect our calculation of the anomalies.
Appendices A and B contain details of calculation of the expansion for the Ricci tensor and the Cotton tensor.
A good starting point in the analysis is to look at the expansion for the Ricci scalar (\ref{a4}). Since the Ricci scalar is
constant for solution of equation (\ref{5}) the subleading terms in the expansion for $R$ should vanish.
In the first sub-leading order, as it is seen from (\ref{a4}),
this gives constraint
\be
\Tr g_{(2)}=-{1\over 2} R(g_{(0)})~~.
\lb{8}
\ee
Note that hereafter we define trace with the help of  metric $g_{(0)}$. Now looking at the
equation (\ref{5}) for components $(\mu\nu)=(ij)$ we find that the  leading term vanishes identically
and the first subleading term vanishes provided constraint (\ref{8}) is taken into account.
Thus, no new constraint on $g_{(2)}$ appears.
Further order terms in the expansion give constraint on higher oder terms $g_{(2k)}$, $k>2$.
At present we are not interested in those terms. A simple relation of this sort comes from the
component $(\mu\nu)=(rr)$ of equation (\ref{5})
\be
\Tr g_{(4)}={1\over 4}\Tr g^2_{(2)}-\beta \epsilon^{ij}\nabla_i\nabla_k g^k_{(2)j}
\lb{9}
\ee
and indicates that (\ref{7}) is not a "total square" as it happened to be in the case of pure GR  \cite{SS}.
The most important constraint comes from components $(\mu\nu)=(r,i)$ of the equation (\ref{5}).
As it follows from (\ref{a4}) and (\ref{b3}) we have that
\be
-\nabla_j g^j_{(2)i}+\partial_i\Tr g_{(2)}+\beta [\epsilon_i^{\ j}(-\nabla_k g^k_{(2)j}+\partial_j \Tr g_{(2)})]=0~~.
\lb{10}
\ee
This can be represented in the form
\be
\nabla_jt^j_{\ i}={\beta\over 2}\epsilon_i^{\ j}\partial_j \Tr g_{(2)}~~,
\lb{11}
\ee
where we have introduced symmetric tensor
\be
t_{ij}=g^{(2)}_{ij}-g^{(0)}_{ij}\Tr g_{(2)}+{\beta\over 2}(\epsilon_i^{\ k}g_{(2)kj}+\epsilon_j^{\ k}g_{(2)ki})~~.
\lb{12}
\ee
Equations (\ref{8}) and (\ref{10}) are the only restrictions on coefficient $g^{(2)}_{ij}$.
Obviously, we can not redefine $t_{ij}$, provided it remains symmetric, to include the right hand side
of (\ref{11}) so that $t_{ij}$ would be covariantly conserved. That it is impossible means that in fact
we  deal with an anomaly. Indeed,
the holographic boundary stress tensor defined as
\be
T_{ij}={1\over 8\pi G_N}t_{ij}
\lb{13}
\ee
has both conformal and gravitational anomalies
\be
\Tr T={l\over 16\pi G_N}~R~,~~\nabla_jT^j_{\ i}=-{\beta \over 32\pi G_N}~\epsilon_{i}^{\ j}\partial_j R~~.
\lb{14}
\ee
When $\beta=0$ the stress tensor defined as in (\ref{13}) agrees with the stress tensor introduced earlier
in  \cite{Balasubramanian:1999re},  \cite{SS},  \cite{CMP}. In particular, this fixes the coefficient in front of
(\ref{13}).  The stress tensor (\ref{13}) also agrees with the one suggested in \cite{Kraus}\footnote{Note that our coupling
$\beta$  differs from the one  used in \cite{Kraus}, exact relation being $\beta=32\pi G_N \beta_{\tt KL}$.}.
We see that the conformal anomaly is not affected by the presence of the Chern-Simons term.
Taking that  conformal symmetry on the boundary of AdS appears as part of bulk diffeomorphisms \cite{Imbimbo:1999bj}
which are broken by the gravitational Chern-Simons term it is rather non-trivial that the conformal anomaly remains unchanged.
On the other hand, the gravitational anomaly (\ref{14}) is entirely due to the Chern-Simons.

The stress tensor $T_{ij}$ is what usually called the metric stress tensor defined as variation of the
action with respect to the metric $g^{ij}$. Since we have at our disposal the objects $h^a_i$ (and their inverse
$h^i_a$) which are "square root" of metric, $g^{ij}=h^i_ah^{aj}$, we can define what might be called
a zweibein stress tensor $T^a_{\ i}$ considering variation of the action with respect to the zweibein
$h^i_a$. Obviously, we have ${\delta \over \delta h^i_a}=2 h^{aj}{\delta\over \delta g^{ij}}$ and hence
$T^a_{\ i}=2h^{aj}T_{ij}$. In Lorentz invariant case antisymmetric part $T^{[a,b]}$, where $T^{ab}=h^{bi}T^a_{\ i}$,
vanishes. For the price of loosing the local Lorentz symmetry the zweibein stress tensor
$T^a_{\ i}$ can be redefined so that the new stress tensor would be
covariantly conserved. Indeed, a new stress tensor
\be
\hat{T}^a_{\ i}=T^a_{\ i}+{\beta l \over 16\pi G_N} \epsilon^a_{\ i}R~,~~\nabla^j \hat{T}^a_{\ j}=0~~
\lb{15}
\ee
is covariantly conserved. However, anomaly does not disappear. It reappears as the local Lorentz anomaly.
Indeed, we have for the new tensor
\be
\epsilon_a^{\ i}\hat{T}^a_{\ i} ={\beta \over 8\pi G_N} R
\lb{16}
\ee
that is clear violation of the local Lorentz symmetry under which  $\delta h^i_a=\delta \phi \epsilon_a^{\ b}h^i_b$.
This is of course well known: the coordinate invariance can be restored for the price of loosing the
local Lorentz invariance.

It is of obvious interest to analyze the gravitational anomaly which may
appear in higher dimension $d=4n+2$ when
gravitational Chern-Simons term\footnote{We mean here the Chern-Simons term
  for the
local Lorentz group $SO(1,d)$. Other possible Chern-Simons
terms, for instance for group $SO(2,d)$, do no seem to produce gravitational
anomaly on the boundary of AdS  \cite{BT}.}
is added to the $(d+1)$-dimensional Einstein-Hilbert action. This is currently under investigation.

\section{Remarks on anomalies in two dimensions}
\setcounter{equation}0

\subsection{Local counterterms, conformal and Lorentz anomalies}
Once the Lorentz symmetry is broken anyway it is allowed to add  local counterterms to the boundary action that are not
Lorentz invariant. Appropriate counterterms depend on the zweibein $h^a_i$, $a=1,2$, rather than on the metric.
It is interesting that by adding such local counterterms we can shift the value of the conformal anomaly--the possibility
which we did not have when dealt only with metric. The counterterm of this sort  was suggested in \cite{Obukhov:1990gy}
\be
I_{\tt ct}={1\over 4}\int d^2x~ h~C^a_{ij}C^{ij}_a~~,
\lb{21}
\ee
where $C^a_{ij}=\partial_i h^a_j-\partial_j h^a_i$ is the anholonomity object for for the zweibein $h^a_i,\ a=1,2$
on the boundary and $h=det~ h^a_i$. Notice, that this term added on the regulated boundary (at fixed value of radial
coordinate $r$) is finite in the limit when $r$ is infinite.

The Lorentz group in two dimensions is abelian so that the Lorentz connection
has only one component
$$
\omega^a_{\ b,i}=\epsilon^a_{\ b}\omega_i~,~~\omega_i={1\over 2}\omega_{ab,i}\epsilon^{ab}~~.
$$
Under local Lorentz and conformal transformation $\delta h^a_i=\delta \sigma h^a_i+\delta \phi \epsilon^a_{\ b}h^b_i$
the Lorentz connection transforms as
\be
\delta \omega_i=\partial_i \delta\phi+\epsilon_i^{\ j}\partial_j \delta \sigma~~.
\lb{2*}
\ee
The counterterm (\ref{21}) changes as follows
\be
\delta I_{\tt ct}={1\over 2} \int d^2x~h~ [\delta\sigma R+\delta \phi K]~~.
\lb{22}
\ee
$R$ is the two-dimensional Ricci scalar which can be expressed in terms of the Hodge dual to the Lorentz connection
one-form
$$
R=2\nabla_i(\tilde{\omega}^i)~,~~\tilde{\omega}^i=\epsilon^i_{\ j}\omega^j~~.
$$
The quantity $K$ that appears in (\ref{22})
has similar expression  in terms of the Lorentz connection itself
\be
K=2\nabla_i(\omega^i)~~.
\lb{23}
\ee
It is invariant under conformal transformations and changes under the local Lorentz transformations.
There is certain similarity between $R$ and $K$ well discussed in \cite{Obukhov:1990gy}.
It is important that both $R$ and $K$ may appear in conformal and/or Lorentz anomaly.
Obviously, adding (\ref{21}) with appropriate coefficient to the boundary effective action
we can always
shift the value of the conformal anomaly and even remove it completely. As a price for that the
quantity $K$ would appear in the Lorentz anomaly.

\subsection{Stress-energy tensor from anomalies}
Two-dimensional black hole can be put on the boundary of three-dimensional anti-de Sitter, the two-dimensional Hawking
effects then would be  encoded in the bulk three-dimensional geometry \cite{SS}.
It is well-known that conformal  anomaly plays important role in two dimensions
and eventually is responsible for the Hawking effect.
Important element in this demonstration \cite{Christensen:1977jc} is the observation that
the conformal anomaly can be integrated to determine the
the covariantly conserved stress-energy tensor. In this subsection we analyze whether this is still
true when the gravitational anomaly is present. Thus, we would like to see whether
 the equations
\be
T_{ij}g^{ij}=a R~,~~\nabla_j T^j_{\ i}=-b \epsilon_i^{\ j}\partial_j R~~,
\lb{24}
\ee
where $a$ and $b$ are some constants, can be integrated and determine the stress energy tensor $T_{ij}$.
Constants $a$ and $b$ can be further related to the central charge in left- and right-moving sectors
of two-dimensional theory as we discuss it in section 5. The exact relation is $a={c_+\over 24\pi}$ and
$b={c_-\over 48\pi}$, $c_\pm=(c_L\pm c_R)/2$.

We start with two-dimensional static metric in the Schwarzschild like form
\be
ds^2=-g(x)dt^2+{1\over g(x)}dx^2~~,
\lb{25}
\ee
where $g(x)$ is some function of the spatial coordinate $x$.
The only non-vanishing Christoffel symbols for this metric are
$$
\Gamma^t_{tx}={g'\over 2g}~,~~\Gamma^x_{tt}={gg'\over 2}~~,
$$
where $g'=\partial_x g$, and the scalar curvature takes the simple form $R=-g''(x)$.
Assuming that components of the stress tensor $T_{ij}$ do not depend on time $t$ we get that  equations (\ref{24})
are equivalent to a set of differential equations
\be
&&T^x_x+T^t_t=-ag'' \nonumber \\
&&\partial_x T^x_x+{g'\over 2g}(T^x_x-T^t_t)=0 \nonumber \\
&&\partial_x T^x_t=bgg'''~~.
\lb{26}
\ee
We chose orientation in which  $\epsilon^{t x}=+1$ when derived (\ref{26}).
These equations can be solved and the solution reads
\be
&&T^t_t=a(-g''+ {g'^2\over 4g}+{C_1\over g}) \nonumber \\
&&T_{xt}=b(g''-{g'^2\over 2g}+{C_2\over g})~~,
\lb{27}
\ee
where $C_1$ and $C_2$ are integration constants. The Hawking
temperature of two-dimensional black hole is $T_H=g'(x_+)/4\pi$, where $x_+$ is location of horizon defined as simple root of function $g(x_+)=0$.
The condition of regularity (see for instance \cite{Solodukhin:1994yz}) of $T^t_t$ and $T_{xt}$ at horizon fixes
the constants $C_1=-{1\over 4}g'^2(x_+)$
and $C_2={1\over 2}g'^2(x_+)$.
The stress tensor thus can be uniquely reproduced from the anomaly equations (\ref{24}).
We see that the gravitational anomaly shows up only in the component $T_{xt}$ which is now non-vanishing
and proportional to $b$. If the two-dimensional space-time is asymptotically flat, i.e. $g(x)\rightarrow 1$
when $x\rightarrow \infty$, then (\ref{27}) describes at infinity a non-vanishing flow
\be
T_{tt}=c_+{\pi\over 6}T^2_H~,~~T_{xt}=c_-{\pi\over 6}T^2_H
\lb{28}
\ee
due to the Hawking particles radiated by black hole.

\section{Black hole entropy from gravitational Chern-Simons term}
\setcounter{equation}0
The gravitational Chern-Simons term is a legitimate action for gravitational field.
It produces covariant field equations which might have sensible solutions\footnote{The corresponding equations of motion
$C_{\mu\nu}=0$  are satisfied for any conformally flat 3d metric.}.
In particular, the constant curvature space-time is always solution of these equations
and remains to be a solution when the gravitational dynamics is governed, as in (\ref{1}), by the sum of
Einstein-Hilbert action and the Chern-Simons term. The BTZ black hole is thus a solution to the equations
(\ref{*}),  as  was first noted in \cite{Kaloper}. On the other hand, it is well known that the expression for the Bekenstein-Hawking entropy
is modified if gravitational action  is   non-linear or even non-local function
of curvature.  In general the entropy is not just a quarter
of horizon area but depends also on the way horizon is embedded in the space-time.
It is thus interesting question whether the gravitational Chern-Simons (\ref{3})  leads to any modifications of the entropy.
The tricky point here is that the Chern-Simons is defined with respect to the Lorentz connection so that one might worry
whether the corresponding entropy is gauge invariant. In this section we analyze this issue.

There are various ways to compute the entropy for a given gravitational action. The most popular
is the Wald's Noether charge method \cite{Wald:1993nt}. It is however a on-shell method which is valid
on the equations of motion. Below we use another method which is universal and does not rely on the equations of motion.
This is the method of conical singularity \cite{Banados:1993qp}, \cite{Solodukhin:1994yz}, \cite{FS}. The idea is to
allow black hole to have
temperature different from the Hawking one. In the Euclidean description this leads to the appearance of  deficit angle
$\delta=2\pi(1-\alpha),~\alpha=T_H/T$ at horizon $\Sigma$. The geometry of manifolds with conical singularities
was analyzed in detail in \cite{FS}.  In particular, it was found that components of the Riemann tensor
contain a singular, delta-function like, part
\be
R^{\alpha\beta}_{\ \ \mu\nu}=(R^{\alpha\beta}_{\ \ \mu\nu})_{\tt reg}
+2\pi (1-\alpha)[(n^\alpha n_\mu)(n^\beta n_\nu)-(n^\alpha n_\nu)(n^\beta n_\mu)]\delta_\Sigma~~,
\lb{31}
\ee
where $(R^{\alpha\beta}_{\ \ \mu\nu})_{\tt reg}$ is non-singular part of the curvature; $(n^\alpha n_\mu)=
n_1^\alpha n^1_\mu+n_2^\alpha n^2_\mu$, $n_1$ and $n_2$ is pair of vectors normal to $\Sigma$ and orthogonal to each other.
Obtained originally in \cite{FS} for static non-rotating metric, this formula was later shown in \cite{MS} to be correct in
the case of  stationary metric.

Taking into account (\ref{31}), the gravitational action in question is now function of $\alpha$. The entropy then is defined
as
\be
S=(\alpha {\partial\over \partial\alpha}-1)|_{\alpha=1}I_{\tt gr}(\alpha)~~.
\lb{32}
\ee
Applying this formula to the Chern-Simons term (\ref{3}) we get\footnote{For spin connection one has that
$\omega=\omega_{\tt reg}+\omega_{\tt sing}$ so that $R_{\tt reg}=d\omega_{\tt reg}+..$ is regular and $R_{\tt sing}=d\omega_{\tt sing}$
is the singular part in (\ref{31}).
Therefore, schematically, one has $\int \omega R=\int \omega_{\tt reg}R_{\tt reg}+2\int \omega_{\tt reg}R_{\tt sing}$.
This gives extra factor of 2 when (\ref{31}) is applied to action (\ref{3}).}
\be
S_{\tt CS}=-{\beta\over 8G_N} \int_\Sigma \omega_{ab,\sigma}h^a_\alpha
h^b_\beta \epsilon^{\mu\nu\sigma}(n^\alpha n_\mu)(n^\beta n_\nu)~~
\lb{33}
\ee
for the entropy.
Note that indices $a,b$ run values from 1 to 3.
In the case of (2+1)-dimensional black hole the horizon $\Sigma$ is circle. Suppose $\varphi$ is the angular coordinate
on $\Sigma$ then vector $\partial_\varphi$ is orthogonal to $n_1$ and $n_2$.
We assume that vector $\partial_\varphi$ together with vector $\partial_\tau$ form a pair of Killing vectors
at horizon. (Outside horizon the Killing vectors are linear combinations of these two vectors.)
 It follows that the integrand in (\ref{33})
is non-vanishing only if index $\sigma=\varphi$. Introducing $\hat{\epsilon}^{\alpha\beta}=\epsilon^{\mu\nu\varphi}(n^\alpha n_\mu)(n^\beta n_\nu)$
expression (\ref{33}) can be re-written as
\be
S_{\tt CS}=-{\beta\over 8 G_N} \int_\Sigma \omega_{ab,\varphi}h^a_\alpha h^b_\beta \hat{\epsilon}^{\alpha\beta}~~.
\lb{34}
\ee
As far as we are aware, the result (\ref{33}), (\ref{34}) for the Chern-Simons entropy  is new.
Under local Lorentz transformations parameterized by $\Omega_{ab}$ the expression (\ref{34})
changes as
\be
\delta S_{\tt CS}=-{\beta\over 8 G_N} \int_\Sigma [\partial_\varphi(\Omega_{ab})\ \hat{\epsilon}^{ab}]\gamma
d\varphi~~,
\lb{35}
\ee
where $\gamma$ is induced measure on $\Sigma$. Since $\partial_\varphi$ is Killing vector the quantity
$\hat{\epsilon}^{ab}=h^a_\alpha h^b_\beta\hat{\epsilon}^{\alpha\beta}$, being considered on $\Sigma$, does not  depend
on $\varphi$.
Therefore, integrating by parts in (\ref{35}) we find that $\delta S_{\tt CS}=0$,
 i.e.  entropy (\ref{34}) is Lorentz invariant in spite the fact
that the Lorentz connection enters explicitly in (\ref{34}).

\bigskip

\bigskip

\noindent{\large\bf The BTZ black hole}

\bigskip

\noindent The BTZ black hole is important, and in fact the only one known, example of black hole in three
dimensions\footnote{For a recent review on the BTZ black hole, conformal field theory and
three-dimensional gravity see \cite{Carlip:2005zn}.}.
Therefore it is interesting to see how our formulas work in this case. The  orthonormal basis
$e^a=h^a_\mu dx^\mu$ for the BTZ metric is
\be
&&e^1=\sqrt{f(r)}dt~,~~e^2={1\over \sqrt{f(r)}}dr~,~~e^3=r (d\varphi+ N(r) dt)~~,
\lb{36}
\ee
where
\be
f(r)={r^2\over l^2}-{j^2\over r^2}-m={(r^2-r^2_+)(r^2-r_-^2)\over l^2r^2}~,~~N(r)=-{j\over r^2}~~.
\lb{3*}
\ee
We have that
\be
m={r^2_++r_-^2\over l^2}~,~~j={r_+r_-\over l}~~.
\lb{3**}
\ee
Here we work in the Lorentzian signature. The analytic continuation to the Euclidean signature was
analyzed in \cite{Carlip:1995qv} and \cite{MS2}.
The vectors orthogonal to horizon are
\be
n_1={1\over \sqrt{f}}(\partial_t-N\partial_\varphi)~,~~n_2=\sqrt{f}\partial_r
\lb{37}
\ee
so that we have that
\be
&&(n^tn_t)=(n^rn_r)=1~,~~(n^\phi n_t)=-N(r_+) \nonumber \\
&&\hat{\epsilon}^{tr}={1\over r_+}~,~~\hat{\epsilon}^{\varphi r}={N(r_+)\over r_+} \nonumber \\
&&\hat{\epsilon}^{12}={1\over r_+}~,~~\hat{\epsilon}^{13}=\hat{\epsilon}^{23}=0~~.
\lb{38}
\ee
Taking into account that  measure $\gamma=r_+$ on $\Sigma$  we find that the expression for entropy takes a simple form
\be
S_{\tt CS}=-{\beta\over 4G_N} \int_0^{2\pi}\omega_{12,\varphi}d\varphi~~.
\lb{39}
\ee
Explicit calculation, making use of eq. (\ref{L}),  shows that
\be
\omega_{12,\varphi}={j\over r_+}~~.
\lb{310}
\ee
The contribution to the entropy due to the Chern-Simons term
\be
S_{\tt CS}=-{\beta\over 4G_N}{2\pi r_-\over l}
\lb{311}
\ee
is thus proportional to the area $2\pi r_-$ of the inner horizon.
That's a curious property of the gravitational Chern-Simons term. Its entropy is apparently
due to degrees of freedom at  inner horizon rather that at the horizon which may be seen by an external observer.
We will comment on this interesting feature later in the paper.
Summing the contributions to the entropy that come from each term in the gravitational action (\ref{1}) the total entropy of BTZ
black hole is
\be
S_{\tt BH}={2\pi r_+\over 4G_N}-{\beta\over l}{2\pi r_-\over 4G_N}~~.
\lb{312}
\ee
Depending on the sign of $\beta$ the contribution of the Chern-Simons term to the entropy may be negative.
This is not a problem as soon as the total entropy (\ref{312}) is positive. This imposes  certain bound on possible values
of $\beta$. We discuss this in the next section.

\section{The boundary CFT calculation}
\setcounter{equation}0
In this section we use the representation when the  Lorentz symmetry  is brocken but the theory is diffeomorphism invarian,
so that the stress-energy tensor of the dual theory is covariantly conserved.
The boundary CFT in question is characterized by different values of the central charge for holomorphic
and anti-holomorphic fields.  The zweibein stress tensor of the theory has both conformal and Lorentz anomalies.
Summarizing our analysis in section 2 we have
\be
h^i_a \hat{T}^a_{\ i}={c_+\over 12\pi}R~,~~\epsilon_a^{\ i}\hat{T}^a_{\ i}={c_-\over 12\pi}R~~,
\lb{41}
\ee
where $c_\pm={1\over 2}(c_L\pm c_R)$ and $c_L$ ($c_R$) is central charge for the left-(right-) moving sector.
Expressions (\ref{14}) and (\ref{16}) give precise values for the central
charge in each sector\footnote{Similar shift in
central charge
was observed in \cite{Blagojevic} within the Brown-Henneaux approach. I thank S. Carlip for drawing my attention to
 this reference.}
\be
c_L={3\over 2}{(l+\beta)\over G_N}~,~~c_R={3\over 2}{(l-\beta)\over G_N}~~.
\lb{42}
\ee
The BTZ black hole corresponds to the sector in the boundary CFT characterized by conformal weights \cite{Brown:1986nw}
\be
h_L={Ml-J\over 2}~,~~h_R={Ml+J\over 2}
\lb{43}
\ee
that are determined by mass $M$ and angular momentum $J$ of black hole. These two parameters are the integrals
$$
M=l\int_0^{2\pi} d\varphi T_{tt}~,~~J=-l\int_0^{2\pi} d\varphi T_{t\varphi}
$$
of the components of the stress tensor defined in (\ref{13}), (\ref{12}). The coefficients in the  Fefferman-Graham
expansion of BTZ metric
are collected in Appendix C.  These are needed for computing the stress tensor using (\ref{13}), (\ref{12}).
We then get
\be
M=M_0-{\beta\over l^2} J_0~,~~J=J_0-\beta M_0~~,
\lb{44}
\ee
where quantities
\be
M_0={r^2_++r^2_-\over 8G_Nl^2}~,~~J_0={r_+r_-\over 4G_Nl}
\lb{45}
\ee
are values of mass and angular momentum in the absence of the Chern-Simons term.
The shift (\ref{44}) has been recently found in \cite{Kraus}. In fact it was known for some time (see  \cite{Blagojevic},
\cite{hehl})
 that  mass and angular momentum in topologically massive gravity are linear combinations of
mass and angular momentum obtained in pure GR.

The entropy in the boundary CFT is computed by the Cardy formula
\be
S_{\tt CFT}=2\pi \left(\sqrt{c_Lh_L\over 6}+\sqrt{c_Rh_R\over 6}~\right)~~.
\lb{46}
\ee
Plugging here the known values for the central charge and conformal weight in each sector we find
\be
S_{\tt CFT}={2\pi r_+\over 4G_N}-{\beta\over l} {2\pi r_-\over 4G_N}
\lb{47}
\ee
that is in perfect agreement with the  black hole entropy (\ref{312}) computed in previous section.
When the bulk theory is pure GR the agreement
is well known \cite{Strominger:1997eq}. The gravitational and CFT entropies still agree when the gravitational
Chern-Simons term is added in the bulk, as we  have just shown. Notice, that this bulk theory   is  much reacher
than GR since it now contains   propagating degrees of freedom.

Apparently, large values (of any sign) of the coupling $\beta$ are not allowed in the theory. There are two
obvious signals of instability for large $\beta$. Central charge in either of two sectors may become negative.
Also, entropy becomes negative when $\beta$ is "too large". These bad things do not happen if parameter
$\beta$ is within the range
\be
|\beta|\leq l~~.
\lb{48}
\ee
This "stability bound" guarantees that both the bulk theory with the gravitational Chern-Simons term and
the boundary CFT with  $c_L\neq c_R$ are well-defined.

\section{Does the Chern-Simons term look deep into black hole?}
\setcounter{equation}0
In theories of gravity involving higher powers of Riemann tensor the black hole entropy is no longer the usual
$A/4G_N$  and is always modified. This is well known and quite well understood.
We refer the reader to  \cite{Jacobson:1993vj} for the Noether charge calculation and to \cite{FS}
for the calculation that uses the conical singularity method. For black holes arising in string theory this issue
was much studied, see review in \cite{LopesCardoso:1999cv}. For BTZ black hole
and higher-dimensional  black holes that reduce to BTZ, this issue was studied in \cite{Saida:1999ec} and recently in
\cite{Kraus:2005vz}. We  would like to discuss here some interesting peculiarities
of  higher curvature modifications of the entropy of BTZ black hole.

The general action of local theory of gravity with higher derivatives can be represented as a power series
in Riemann curvature. This in fact is true also for a non-local theory however each term in such expansion then
would contain non-local factors. Keeping theory local the quadratic term in our action would be something like this
\be
W=\int \left( {a_1\over 24\pi} R^2+{a_2\over 16\pi}R_{\mu\nu}^2+{a_3\over 16\pi}R_{\alpha\beta\mu\nu}^2\right)~~.
\lb{51}
\ee
The corresponding contribution (see  \cite{Jacobson:1993vj}  and \cite{FS} for more detail) to the entropy is
\be
S=-\int_\Sigma \left({a_1\over 3}R+{a_2\over 4}R_{\mu\nu}(n^\mu n^\nu)+{a_3\over 2}R_{\mu\nu\alpha\beta}(n^\mu n^\alpha)(n^\nu n_\beta)
\right)
\lb{52}
\ee
as can be easily obtained using method outlined in section 4. Applying this to BTZ  black hole
we notice that the BTZ metric is locally AdS and hence the Riemann tensor factorizes
$R_{\alpha\beta\mu\nu}={1\over l^2}(G_{\beta\mu}G_{\alpha\nu}-G_{\alpha\mu}G_{\beta\nu})$. This factorization
and that vectors $n_1$ and $n_2$ are orthonormal  lead to an interesting conclusion that nothing
in the integrand in (\ref{52}) depends on the parameters of black hole. Those parameters enter  (\ref{52}) only via
 area of $\Sigma$, i.e. via $r_+$,
\be
S=(a_1+a_2+a_3){2\pi r_+\over l}~~.
\lb{53}
\ee
Obviously this property remains in place when higher powers of  curvature are included in the action. In fact we can state that
any local theory of gravity that is non-linear in curvature results in the entropy
which takes the form
\be
S_{\tt non}= \mu(a_i, l) {2\pi r_+\over l}~~,
\lb{54}
\ee
where $\mu(a_i, l)$ is some function of  higher curvature couplings $a_i$
and the AdS scale $l$ but not of the parameters of black hole.
Similar result was recently derived in \cite{Kraus:2005vz}.
The higher derivative theory of gravity, provided it is formulated
in terms of gauge invariant objects, i.e. the Riemann tensor, thus sees only the radius $r_+$ of outer horizon
of BTZ black hole and leaves $r_-$ unnoticed\footnote{In general, this may be different in the case of non-local theory of gravity.
Such a theory may produce logarithmic terms in the entropy and both $\ln r_+$ and $\ln r_-$ are a priori possible.
The concrete calculation in \cite{MS2} however shows that to the leading order such entropy is determined by $r_+$ only,
$r_-$ appearing in the subleading terms.}.

The gravitational Chern-Simons term, as we have seen in section 5, shows radically different behavior. Its entropy is
proportional to the area $2\pi r_-$ of  inner horizon so that it is $r_+$ that is now  unnoticed.
This is despite the fact that  the entropy is actually given by integral (\ref{33}) over  outer horizon.
This is an interesting feature of the gravitational
Chern-Simons term that it seems to see the interior of the black hole. The Lorentz connection apparently does the trick.
Most dramatically this feature manifests itself when the gravitational action contains the Chern-Simons term only.
The BTZ metric is still a solution to the field equation. Its temperature, mass and angular momentum are non-vanishing
and hence, thermodynamically, there must be some  entropy and that entropy is precisely $S_{\tt CS}$ defined in (\ref{311})
and determined by the area of  inner horizon. Notice, that it constitutes entire entropy of black hole in this case!
This observation poses interesting questions that may be challenging to  our present understanding of the black hole entropy.
The obvious one is whether there should  be some degrees of freedom associated with the inner (rather than with the outer)
horizon which would be responsible for this entropy? We leave this and other questions for the future.

\bigskip

\bigskip

\noindent {\large \bf Acknowledgments}

\bigskip

\noindent I would like to thank K. Krasnov for a helpful remark.
This work is supported in part by  DFG grant Schu 1250/3-1.

\appendix{Curvature components and their expansion}
\setcounter{equation}0

Components of the Riemann tensor are
\be
R^r_{\ irj}&=&{1\over 2}[-g''+{1\over 2}g'g^{-1}g']_{ij} \nonumber \\
R^r_{\ ikj}&=&-{1\over 2}[\nabla_kg'_{ij}-\nabla_jg'_{ik}] \nonumber \\
R^{l}_{\ ikj}&=&R^{l}_{\ ikj}(g)-{1\over 4}g'_{ij}g^{ln}g'_{nk}+{1\over 4}g'_{ik}g^{ln}g'_{nj}~~,
\lb{a1}
\ee
where $g'\equiv \partial_r g$.
Components of Ricci tensor are
\be
R_{ij}&=&R_{ij}(g)-{1\over 2} g''_{ij}-{1\over 4}g'_{ij}\Tr(g^{-1}g')+{1\over 2}(g'g^{-1}g')_{ij} \nonumber \\
R_{ri}&=&{1\over 2}[\nabla^k(g^{-1}g')_{ki}-\nabla_i\Tr(g^{-1}g')] \nonumber \\
R_{rr}&=&-{1\over 2}\Tr(g^{-1}g'')+{1\over 4}\Tr(g^{-1}g'g^{-1}g')
\lb{a2}
\ee
and the Ricci scalar is
\be
R=R(g)-\Tr(g^{-1}g'')-{1\over 4}[\Tr(g^{-1}g')]^2+{3\over 4}\Tr(g^{-1}g'g^{-1}g')~~.
\lb{a3}
\ee
The leading terms in the Fefferman-Graham expansion of the curvature tensors are
\be
R_{ri}&=&[-\nabla_ng^n_{(2)i}+\partial_i\Tr g_{(2)}]e^{-2r}+... \nonumber \\
R^k_{\ i}&=&-2\delta^k_i +[R^k_i(g_{(0)})+\delta^k_i \Tr g_{(2)}]e^{-2r}+... \nonumber \\
R_{rr}&=&-2+[-4\Tr g_{(4)}+\Tr g^2_{(2)}]e^{-4r}+... \nonumber \\
R&=&-6+[R(g_{(0)})+2\Tr g_{(2)}]e^{-2r}+...
\lb{a4}
\ee
For the constant curvature $R=-6$ metric we have a constraint
\be
\Tr g_{(2)}=-{1\over 2}R_{(0)}~~.
\lb{a5}
\ee

\appendix{Components of the Cotton tensor and their expansion}
\setcounter{equation}0

In space-time with constant Ricci scalar $R=-6$ the Cotton tensor is defined as
\be
C_{\alpha\beta}=\epsilon_{\alpha}^{\ \mu\nu}\nabla_\mu R_{\nu\beta}~~.
\lb{b1}
\ee
For the Levi-Civita symbol we have that $\epsilon^{rij}=\epsilon^{ij}$ where $\epsilon^{ij}$ is defined
for the 2d metric $g_{ij}(r,x)$.

In terms of $g_{ij}(r,x)$ we get for the components of (\ref{b1})
\be
C_{ri}&=&-\epsilon_n^{\ k}\nabla_k R^n_i+{1\over 2}\epsilon^{kn}g'_{ki}R_{rn} \nonumber \\
C_{rr}&=&\epsilon^{ij}[\nabla_i R_{rj}-{1\over 2}(g^{-1} g')^k_i R_{kj}] \nonumber \\
C_{ij}&=&-\epsilon_i^{\ k}[\partial_r R_{kj}-{1\over 2}(g^{-1}g')^n_jR_{kn} -\nabla_k R_{rj}-{1\over 2}g'_{kj}R_{rr}]~~.
\lb{b2}
\ee
Taking into account the constraint (\ref{a5}) we find the following expansion for the components of the
Cotton tensor
\be
C_{ri}&=&\epsilon_i^{\ j}[-\nabla_k g^k_{(2)j}+\partial_j\Tr g_{(2)}]\ e^{-2r}+... \nonumber \\
C_{rr}&=&-\epsilon^{ij}\nabla_i\nabla_k g^k_{(2)j} \ e^{-4r}+... \nonumber \\
C_{ij}&=&0+O(e^{-2r})~~,
\lb{b3}
\ee
where  the leading term (of order $e^{0r}$) in the expansion of $C_{ij}$ vanishes due to  constraint (\ref{a5}).

\appendix{The BTZ metric in normal coordinates}
\setcounter{equation}0

The BTZ metric can be brought to the normal coordinates in the form (\ref{6}) as follows
\be
&&ds^2=dr^2-({r^2_+\over l^2}\sinh^2{r\over l}-{r^2_-\over l^2}\cosh^2{r\over l})dt^2 \nonumber \\
&&+({r^2_+\over l^2}\cosh^2{r\over l}
-{r^2_-\over l^2}\sinh^2{r\over l})d\varphi^2-{2r_+r_-\over l}dtd\varphi~~,
\lb{c1}
\ee
where $r_+$ ($r_-$) is radius of outer (inner) horizon.
The coefficients in the expansion (\ref{7}) of this metric are
\be
g^{(0)}_{tt}&=&-{1\over l^2}g^{(0)}_{\varphi\varphi}=-{(r^2_+-r^2_-)\over 4l^2} \nonumber \\
g^{(2)}_{tt}&=&{1\over l^2}g^{(2)}_{\varphi\varphi}={(r^2_++r^2_-)\over 2l^2}~,~~g^{(2)}_{t\varphi}=-{r_+r_-\over l}~~.
\lb{c2}
\ee
We choose orientation in which  $\epsilon_{t}^{\ \varphi}=-1/l$.

\end{document}